\documentclass[conference]{IEEEtran}
\IEEEoverridecommandlockouts
\usepackage{cite}
\usepackage{amsmath,amssymb,amsfonts}
\usepackage{algorithmic}
\usepackage{graphicx}
\usepackage{textcomp}
\usepackage{xcolor}
\usepackage{caption}
\usepackage{multirow}
\usepackage{subcaption}
\usepackage{xcolor,colortbl}
\usepackage{makecell}
\usepackage{booktabs}
\usepackage{url}
\usepackage{hyperref}
\def\BibTeX{{\rm B\kern-.05em{\sc i\kern-.025em b}\kern-.08em
    T\kern-.1667em\lower.7ex\hbox{E}\kern-.125emX}}
\begin{document}

% \title{On the Impact of  Visuospatial Representation and Ability on Code Comprehension}
\title{From RSSE to BotSE: Potentials and Challenges Revisited after 15 Years}

\author{
\IEEEauthorblockN{Walid Maalej}
\IEEEauthorblockA{\textit{Applied Software Technology} \\
\textit{Universität Hamburg}\\
Hamburg, Germany \\
walid.maalej@uni-hamburg.de
}
}
\maketitle

\begin{abstract}
Both recommender systems and bots should proactively and smartly answer the questions of software developers or other project stakeholders to assist them in performing their tasks more efficiently. 
This paper reflects on the achievements from the more mature area of Recommendation Systems in Software Engineering (RSSE) as well as the rising area of Bots in Software Engineering (BotSE). 
We discuss the similarities and differences, briefly review current state of the art, and highlight three particular areas, in which the full potential is yet to be tapped: a more socio-technical context awareness, assisting knowledge sharing in addition to knowledge access, as well as covering repetitive or stimulative scenarios related to requirements and user-developer interaction.

\end{abstract}


\section{Introduction}

Software engineering (SE) was among the first fields to research  and adopt recommendation systems (or recommenders) when they emerged in the early 2000s \cite{Ricci:RecSys:2010, RSSE:08, Robillard:RSSE:14}.  
At the time, SE researchers presented tools like CodeBroker in 2002 \cite{Yunwen:ASE:02} to recommend methods for reuse, Hipikat in 2005 \cite{Cubranic:TSE:05} to recommend relevant related artifacts, or Mylyn \cite{Kersten:FSE:06} in 2006 to recommend the code elements that are relevant for the current task. 
Recommendation  approaches quickly made their way into software engineering practice, perhaps most prominently in form of code auto-completion and recommendation features\footnote{\url{https://geekflare.com/best-ai-powered-code-completion-tools/}}, which have become indispensable in modern Integrated Development Environments (IDEs). 
%Workshop series and a book.

Similarly, about a decade later \cite{Storey:FSE:16}, the emerging technology of bots (or chatbots) has been researched and applied in SE context---long before the recent hype around Large Language Models and Generative AI, which has brought this technology in the meantime into almost every domain \cite{Bubeck:arXiv:23}. 
A repository that emerged from the BotSE workshop series \cite{Botse:19} collecting papers on software bots  \cite{Bot_Research_Repository} includes dozens of paper on this subject. 
It is more and more common for developers to interact with bots in their daily work, e.g.~to solve merge conflicts or to highlight and solve dependability issues. Wyrich et al.~\cite{Wyrich:BotSE:21} estimate that by 2021 about one third of pull requests in GitHub are created by software bots.

While these two technologies are apparently different and the communities largely distinct, for SE, both recommenders and bots share the same goal: that is supporting developers and project stakeholders ideally in a smart and proactive way to address their information needs and automate parts of their tasks. 
Therefore, it seems obvious to reflect on both areas, their potentials, and challenges. 
15 years ago, with the rise of RSSEs, we published a short paper on the potentials and challenges of recommendation systems for software engineering \cite{Happel:RSSE:2008}.
In this invited position paper, we revisit our work from 2008 to reflect on what has been achieved and map the work to bots. 
We highlight main challenges that we think are still open along with the potential the technologies still bear. 

\section{Recommenders versus Bots}
There are similarities and differences between recommenders and bots. 
On the one hand, both technologies aim at \textbf{assisting} users in performing their tasks. 
In the context of SE, those users are developers and project stakeholders like requirements analysts, testers, managers, or domain experts -- all performing SE work. 
The overall goal is to increase the productivity and effectiveness of a certain task, either by answering questions that might emerge to the stakeholders or by increasing the overall degree of automation \cite{Robillard:RSSE:14, Storey:FSE:16}. 

Moreover, some kind of \textbf{``smartness''} is expected from both technologies. 
In the case of recommenders, usually a retrieval and processing of information as well as a \textit{ranking} of items will take place before the recommendation is shown. 
In the case of bots, at least a mimicking and  \textit{understanding} of the user prompts and the entire conversation will take place, sometimes also including a \textit{reasoning} \cite{Bubeck:arXiv:23, Robe:FSE:22}, e.g.~about related situations or the overall context.  
Usually, both technologies require a human (i.e.~a developer or another stakeholder) to check their outcome before the final decision is made. They are designed to support human decision making and not replace it (i.e.~human in the loop).

On the other hand, there are differences related to what bots and recommenders do and how they do things. 
Concerning the \textbf{``What"}, bots do not necessarily follow the goal to recommend items. 
Although, one might intuitively expect from a software bot to process, select, and rank information (as recommenders do), bots often conduct simpler tasks, e.g.~translate a natural language query into a command, run a script and report the results, summarize a discussion, or just inform a developer about a new or a resolved issue.
Modern bots as GitHub Copilot \cite{Nguyen:MSR:22} may combine multiple aspects and tasks into a single  conversation, acting as an integrated interface to multiple complex services \cite{Lebeuf:BotSE:19}. 
Recommenders, however, focus on a specific type of recommendation items.   

Concerning the \textbf{``How''}, while recommendations are integrated into the user interface of stakeholders' tools (e.g.~in form of a search bar, a tooltip, or an auto-complete widget), bots often present a different, more human-like interaction interface (and style) with users. 
This is often referred to as conversational user interface (thus the designation conversational bots). 
This perhaps explains why questions related to human- and social- factors are among the most important when it comes to BotSE research \cite{Lebeuf:BotSE:19,Storey:FSE:16,Storey:Dagstuhl:20}.
Such questions include the design of bot personas, potential biases or harms a bot might expose, and the impact on stakeholders' affects, privacy, or agency. 
The smartness of bots does not only (and not necessarily) cover the understanding, reasoning, and prediction of technical questions at hand---but also understanding and interacting with the human (developer or stakeholder). 

Despite these principal differences, in the domain of SE (which often aims at increasing automation) it is expected that bots will more and more integrate recommendation engines and systems, as it is, e.g., the case of GitHub Copilot \cite{Nguyen:MSR:22}. Santhanam et al.~\cite{santhanam:PeerJ:22} recently observed: ``Research on bots in software engineering began by automating tasks through scripts [...], but with time the field has become much more interdisciplinary. A fully-fledged development process consists of designing the bot (persona, behavior, mode of interaction), and implementing their abilities (training of machine learning models, development and deployment)." 
The smartness of bots for SE is and will continue to cover more technical aspects in addition to human aspects.

%and automation is key to software researchers and tool vendors, BotSE would, very likely, present or integrate some sort of RSSEs. 

\section{RSSE Potentials and Challenges Revisited}
Nowadays recommendation systems certainly have a higher maturity compared to when they were introduced two decades ago. 
Auto-completion, answer retrieval from software repositories and Q\&A sites, code examples recommendations, defect prediction, or duplicate bug reports predictions are just a few examples of well-researched RSSE approaches, which have made their way to modern tools like IDEs, Code Repositories, Build and Test Infrastructures, or Issue Trackers.

In 2008, we noted \cite{Happel:RSSE:2008} five main limitations to the few (by then)  available RSSE approaches:
\begin{itemize}
    \item By then, limited use cases (bug fixing and coding). By now, research has certainly overcome this limitation \cite{Robillard:RSSE:14}. RSSE research and tools span the whole software lifecycle from requirements to deployment and evolution as a recent study shows \cite{Savary:SPE:22}. Still, the same study also shows that most recommenders cover Code/API search, completion, repair and understanding \cite{Savary:SPE:22}.   
    Current and future focus will likely be on the optimization of the performance and integration of recommendations into the workflows and tools.
    
    \item By then, limited knowledge representation and hard-coded heuristics (e.g.~in Mylyn or Hapikat). 
    By now, this limitation has become less apparent as much follow-up research in the last decade has used machine learning approaches, learning from existing software project data, as well as from available web resources such as Wikipedia and StackOverflow \cite{Savary:SPE:22} to recommend items (e.g. bug assignments or code reviewers from previous resolutions, requirements priorities from user feedback items, or fixes and code changes from Q\&A sites). 
    Language models as BERT or GPT are recently becoming a main enabler of predictions and recommendations. Nevertheless, heuristics and rules are still important to represent RSSE knowledge as Savary et al.~observed \cite{Savary:SPE:22}.
   
    \item By then, centralized and inflexible architectures. 
    By now, and with the maturity of RSSEs research and their  adoption in practice,  more and more scalable and flexible approaches have been discussed in literature and made available to practitioners, often fully integrated (and hardly noticeable) into developers' tool such as IDEs, issue trackers, or continuous build infrastructures. 
    Savary et al.~\cite{Savary:SPE:22}~found about 60\% off surveyed software assistants use a local knowledge source (as the browser history and execution environment) while $\sim$40\% use a remote shared model/knowledge repository (as a search engine or a documentation repository).   
    
    \item By then, limited context-awareness. 
    By now, we believe that there has been only a limited progress in this area. 
    We think that context is still barely considered when recommending items, and if, then usually only a certain type of context (e.g.~the task at hand, \textit{or} the interaction history, \textit{or} the experience of the developer, \textit{or} the conversation etc.). 
    An in-depth observation and modeling of context including the personal and organizational preferences of stakeholders, the priority and nature of work at hand, the available/missing knowledge, are key to delivering intelligent personalized support in software engineering work. 
    This remains an open challenge for future research.
    
    \item By then, only information provision was supported.    
    But, developers and other stakeholders can be supported not only to access information needed but also to \textit{share} information with others who may need it \cite{Happel:RSSE:2008}. 
    By now, this challenge still holds.  
    As the vast majority of work is on supporting information access, we think  this limitation still holds, limiting the potentials of RSSEs. 
    After two decades of research and development, most RSSEs rely on information explicitly available in a certain software repository (e.g.~in Git or in StackOverflow) but not on the information in the heads of developers and other stakeholders such as users and experts. 
    Recommending for people with certain knowledge ``in mind'' to share their knowledge in a specific situation with a specific person is still a rarely researched area. 
\end{itemize}

\section{The Bots Perspective}
To get a preliminary overview on current state of the art and initiate a comparison with the RSSE perspective, we conducted a brief preliminary analysis of the papers collected by the BotSE community \cite{Bot_Research_Repository}. 
Although this is not a systematically collected dataset, it certainly represents a good starting point for research trends and community focus. 
The BotSE workshop series can be considered an anchor around current research on bots in software engineering context.

\begin{figure}[]
    \centering    \includegraphics[scale=0.55]{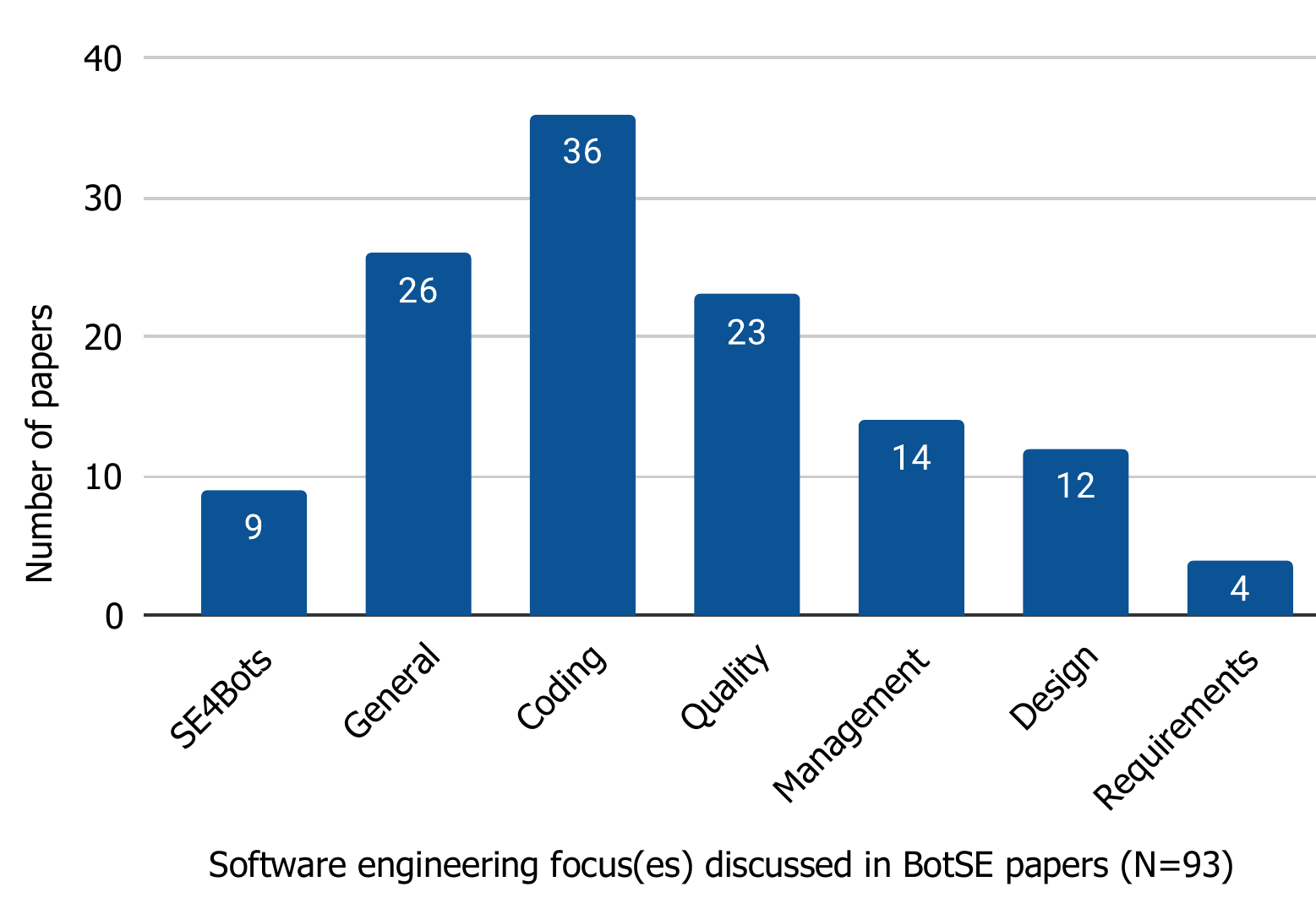}
    \caption{Software engineering activities supported in the BotSE literature  \cite{Bot_Research_Repository}. One papers can cover multiple activities.}.
    \label{fig:botse-focus}
\end{figure}

We counted 117 entries in the repository as of March 2023. 
These have appeared in diverse software engineering venues in addition to the BotSE workshops, including ICSE, TSE, and ASE. 
We found three duplicate entries and 21 papers that did not have a software engineering focus (also in the broad sense): leaving 93 papers which we labeled according to our RSSE landscape introduced in 2008 \cite{Happel:RSSE:2008}. That is, we checked: 

\begin{itemize}
    \item What software engineering activities and focuses are discussed in the BotSE paper?
    \item If the paper is about bots to assist SE (not SE for bots):
    \begin{itemize}
        \item What is the recommended item (i.e.~technical or collaboration recommendation)?
        \item When does the bot make the recommendation (i.e. to access information or to share information)?
   \end{itemize}
\end{itemize}

\begin{figure}
    \centering   \includegraphics[scale=0.55]{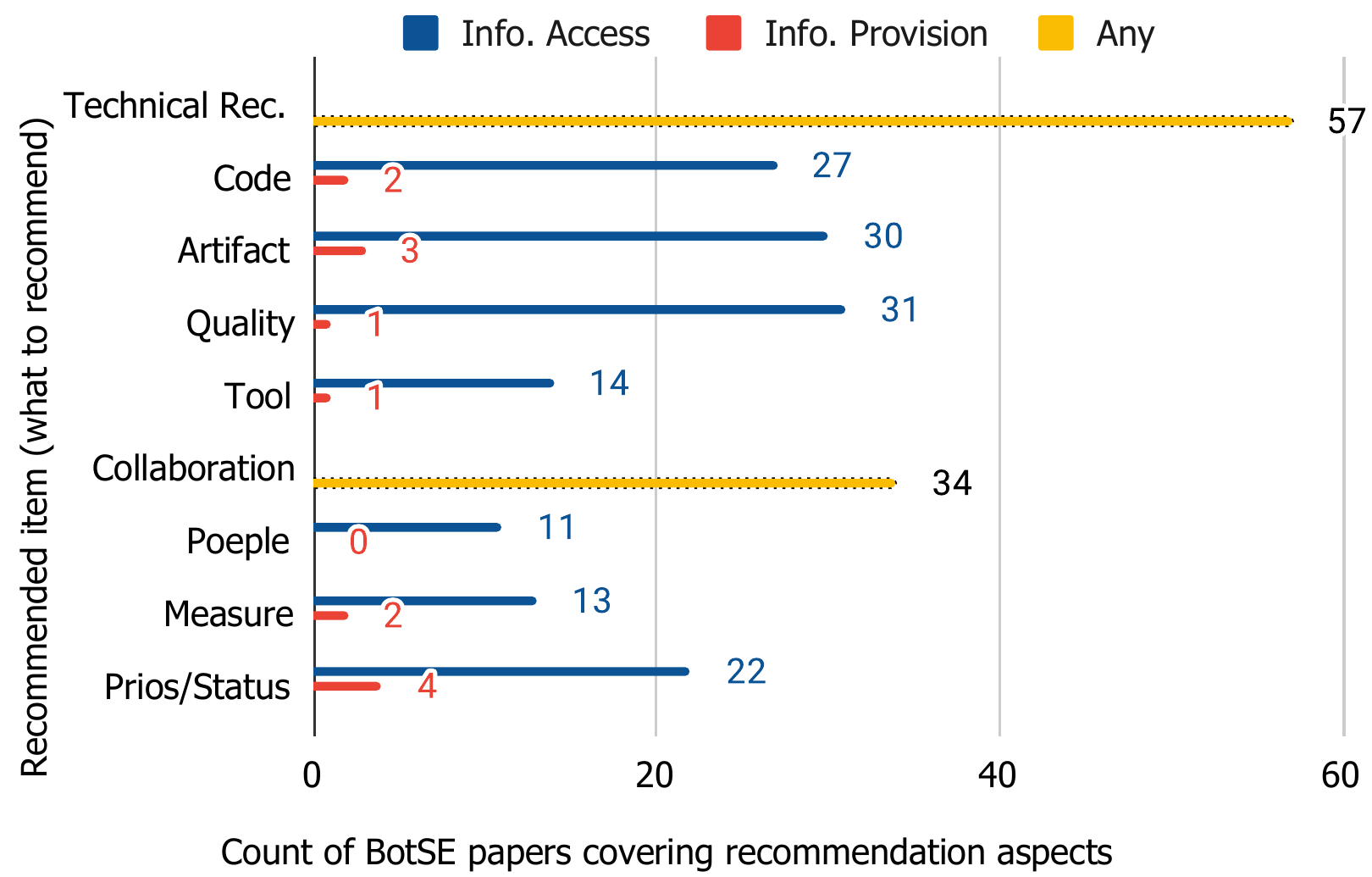}
    \caption{What is recommended and when (to access/share information) \cite{Happel:RSSE:2008}, revisited in the BotSE literature\cite{Bot_Research_Repository}.}
    \label{fig:botse-what-when}
\end{figure}

The results are shown on Figure \ref{fig:botse-focus} and Figure \ref{fig:botse-what-when}.
Nine out of 93 relevant papers are about software engineering \textit{for} bots, e.g.~how to design, develop, and test bots -- a rather small portion of the research effort (at least in this dataset). 
The rest cover a broad spectrum of SE scenarios and activities. 
The most popular scenarios supported by bots (36 papers) are related to coding (such as commit-, refactoring-, or code questions- bots as the recent RSSE review also found \cite{Savary:SPE:22}), followed by quality measures bots (23 papers), and then by management (14) and design/architecture (12). 
We found only 4 papers discussing bots related to requirements.
This is rather surprising given the huge advance in Natural Language Processing for Requirements Engineering \cite{Liping:CS:21}, the large need for assistants and automation, as well as the potential of recommendation technologies for requirements work \cite{Felfernig:2013}. 

Overall, these results are understandable as programming, code analysis, and defect prediction/repair are among the most atomic, concise, and actionable tasks to partly automate. 
Moreover, most bots supporting these tasks build on social development environments such as GitHub, StackOverflow, or Build Infrastructures, where developers are used to get notified, interact, comment, ask for clarifications, and make small incremental steps. 
Still, we think that management, design, and particularly requirements work bear much potential for feedback, awareness, and support by bots.
For instance, \textbf{developer-user interactions}, e.g.~in app stores or in issue trackers, are gaining more and more importance and have been shown to have an impact on the product quality and project success \cite{hassan:emse:2018,Martens:RE:19}.
Bots could play a major role to scale such interaction to all users \cite{Martens:RE:19}.

The results of ``what to recommend and when'', which are depicted on Figure \ref{fig:botse-what-when}, are also interesting. 
The large majority of bots still focus on suggesting technical items like source code, technical artifacts as API documentation or model elements, patches, and SE tools to use. 
As bots particularly cover the conversational and social aspects, it remains interesting (and worth investigating further) whether the full potential of bots with regard to assisting  collaborative aspects is sufficiently covered in current research. 
For instance, bots to recommend other people or collaboration measures seem rather rare.

As for the ``when to assist/recommend'' the picture is much more clear. 
Scenarios of bots supporting the knowledge sharing seem barely discussed in the BotSE literature. 
And if it is the case, then rather at a meta level \cite{Storey:FSE:16} or as a side aspect in addition to the support of information access. 
One recent paper by Josephs et al.~suggested a bot for capturing design decisions from developer conversations \cite{Josephs:ICSA:22}. 
Another, by Robe et al.~\cite{Robe:FSE:22}, suggested and studied knowledge exchange in 
pair programming conversations with bots. 
Finally,  Phaithoon et al.~\cite{Phaithoon:ASE:22}, discussed a bot for self-admitted technical debt, which also includes sharing knowledge about design and priorities. 
We think that these few works only scratch the surface of the actual potential of bots to mediate knowledge exchange and tighten knowledge sharing among stakeholders in SE context.

\section{Discussion and Summary}
While dephased by a decade and despite the rather independent research efforts, the fields of RSSE and BotSE are closely related.  
It is sometimes even hard to draw the exact line in-between as the definitions of the terms bot and recommender might be challenging \cite{Lebeuf:BotSE:19, Storey:Dagstuhl:20} or used interchangeably \cite{Savary:SPE:22}.
A bot might include one or more recommendations while a recommender might also process its data or get triggered by a bot. 
We think that, in future, most SE bots will go beyond simple ``dumb'' bots that runs scripts, to include predictions, reasoning, item rankings, and recommendations. 
This might help overcome the limitation of bot acceptance by developers \cite{Wyrich:BotSE:21}. 
After all, the value of a RSSE or a BotSE should be in the \textit{usefulness} of the suggestion much more than the accuracy, completeness, or optimization degree.

It is important to continue discussing whether and if yes how RSSEs differs from BotSEs. 
We tried to take a first step to reflect on the potentials and challenges from both areas. 
A more systematic approach is needed for better conclusions. 
Clearly, there are more aspects which should be considered when comparing both fields and distilling the learnings, such as: how similar are the bots and RSSE solutions and architectures? Is BotSE research heading the same direction as RSSE? Is it necessarily for a better acceptance of software bots that they include recommendations? Can the empirical evidence and evaluations collected from RSSEs be applied to bots? 
Success and failure stories of RSSEs (and botSE) should be analyzed in details and learning  derived. 

From our preliminary reflection, we observed the following:

\begin{itemize}
    \item BotSEs and RSSEs research should continue investigate how to model and use the socio-technical context of developers and stakeholders, to deliver a more useful, precise, effective, and personalized assistance.
    \item BotSEs and RSSEs can be used to tighten knowledge sharing in SE. After 15 years, this area is still barely researched, its potential barely tapped.  
%    \item Human and social aspects, the exact wording when interacting with developers, to ensure positive feedback and positive affect.
    \item Bots are able to scale certain interactions and tasks, which used to have a rather low priority to software projects but are becoming more and more important:  particularly the interaction with end users or social interactions and feedback with developers.  
%    \item easier feedback from the developers through conversation. More accurate recommendations. 
\end{itemize}

Bots and recommenders came to SE to stay. It will be again  interesting to revisit their potential and challenges in 15 years.

% integration into the workflow/tool is a key for success (see mylyn or code completion tools). But also switch-off and configurability (e.g. to switch off).

%    This creates additional challenges such as potential bias, making up false but convincing arguments etc.

\section*{Acknowledgment} We thank the BotSE'23 workshop organizers for the invite and comments; Mathieu Nassif for his feedback; and Christian Rahe for helping with labeling part of the papers.

%\IEEEtriggeratref{12}
\bibliographystyle{IEEEtran}

\begin{thebibliography}{10}
\providecommand{\url}[1]{#1}
\csname url@samestyle\endcsname
\providecommand{\newblock}{\relax}
\providecommand{\bibinfo}[2]{#2}
\providecommand{\BIBentrySTDinterwordspacing}{\spaceskip=0pt\relax}
\providecommand{\BIBentryALTinterwordstretchfactor}{4}
\providecommand{\BIBentryALTinterwordspacing}{\spaceskip=\fontdimen2\font plus
\BIBentryALTinterwordstretchfactor\fontdimen3\font minus
  \fontdimen4\font\relax}
\providecommand{\BIBforeignlanguage}[2]{{%
\expandafter\ifx\csname l@#1\endcsname\relax
\typeout{** WARNING: IEEEtran.bst: No hyphenation pattern has been}%
\typeout{** loaded for the language `#1'. Using the pattern for}%
\typeout{** the default language instead.}%
\else
\language=\csname l@#1\endcsname
\fi
#2}}
\providecommand{\BIBdecl}{\relax}
\BIBdecl

\bibitem{Ricci:RecSys:2010}
F.~Ricci, L.~Rokach, B.~Shapira, and P.~B. Kantor, \emph{Recommender Systems
  Handbook}.\hskip 1em plus 0.5em minus 0.4em\relax Springer New York, NY,
  2010.

\bibitem{RSSE:08}
\emph{Proceedings of the 2008 International Workshop on Recommendation Systems
  for Software Engineering, {RSSE} 2008, Atlanta, GA, USA, November 9,
  2008}.\hskip 1em plus 0.5em minus 0.4em\relax {ACM}, 2008.

\bibitem{Robillard:RSSE:14}
M.~P. Robillard, W.~Maalej, R.~J. Walker, and T.~Zimmermann,
  \emph{Recommendation Systems in Software Engineering}.\hskip 1em plus 0.5em
  minus 0.4em\relax Springer Publishing Company, Incorporated, 2014.

\bibitem{Yunwen:ASE:02}
Y.~Ye and G.~Fischer, ``Supporting reuse by delivering task-relevant and
  personalized information,'' in \emph{Proceedings of the 24th International
  Conference on Software Engineering}.\hskip 1em plus 0.5em minus 0.4em\relax
  ACM, 2002, p. 513–523.

\bibitem{Cubranic:TSE:05}
D.~Cubranic, G.~C. Murphy, J.~Singer, and K.~S. Booth, ``Hipikat: A project
  memory for software development,'' \emph{IEEE Transactions on Software
  Engineering}, vol.~31, no.~6, pp. 446--465, 2005.

\bibitem{Kersten:FSE:06}
M.~Kersten and G.~C. Murphy, ``Using task context to improve programmer
  productivity,'' in \emph{Proceedings of the 14th ACM SIGSOFT international
  symposium on Foundations of software engineering}, 2006.

\bibitem{Storey:FSE:16}
M.-A. Storey and A.~Zagalsky, ``Disrupting developer productivity one bot at a
  time,'' in \emph{Proceedings of the 2016 24th ACM SIGSOFT International
  Symposium on Foundations of Software Engineering}, 2016.

\bibitem{Bubeck:arXiv:23}
S.~Bubeck, V.~Chandrasekaran, R.~Eldan, J.~Gehrke, E.~Horvitz, E.~Kamar,
  P.~Lee, Y.~T. Lee, Y.~Li, S.~Lundberg \emph{et~al.}, ``Sparks of artificial
  general intelligence: Early experiments with gpt-4,'' \emph{arXiv preprint
  arXiv:2303.12712}, 2023.

\bibitem{Botse:19}
E.~Shihab and S.~Wagner, Eds., \emph{Proceedings of the 1st International
  Workshop on Bots in Software Engineering, BotSE@ICSE 2019, Montreal, QC,
  Canada, May 28, 2019}.\hskip 1em plus 0.5em minus 0.4em\relax {IEEE} / {ACM},
  2019.

\bibitem{Bot_Research_Repository}
A.~Abdellatif, B.~Khaled, and E.~Shihab, ``A repository of research articles on
  software bots,'' http://papers.botse.org.

\bibitem{Wyrich:BotSE:21}
M.~Wyrich, R.~Ghit, T.~Haller, and C.~M{\"u}ller, ``Bots don’t mind waiting,
  do they? comparing the interaction with automatically and manually created
  pull requests,'' in \emph{2021 IEEE/ACM Third International Workshop on Bots
  in Software Engineering (BotSE)}.\hskip 1em plus 0.5em minus 0.4em\relax
  IEEE, 2021, pp. 6--10.

\bibitem{Happel:RSSE:2008}
H.-J. Happel and W.~Maalej, ``Potentials and challenges of recommendation
  systems for software development,'' in \emph{Proceedings of the 2008
  International Workshop on Recommendation Systems for Software
  Engineering}.\hskip 1em plus 0.5em minus 0.4em\relax ACM, 2008, p. 11–15.

\bibitem{Robe:FSE:22}
P.~Robe, S.~K. Kuttal, J.~AuBuchon, and J.~Hart, ``Pair programming
  conversations with agents vs. developers: Challenges and opportunities for se
  community,'' in \emph{Proceedings of the 30th ACM Joint European Software
  Engineering Conference and Symposium on the Foundations of Software
  Engineering}, ser. ESEC/FSE 2022.\hskip 1em plus 0.5em minus 0.4em\relax ACM,
  2022, p. 319–331.

\bibitem{Nguyen:MSR:22}
N.~Nguyen and S.~Nadi, ``An empirical evaluation of github copilot's code
  suggestions,'' in \emph{Proceedings of the 19th International Conference on
  Mining Software Repositories}.\hskip 1em plus 0.5em minus 0.4em\relax ACM,
  2022, p. 1–5.

\bibitem{Lebeuf:BotSE:19}
C.~Lebeuf, A.~Zagalsky, M.~Foucault, and M.-A. Storey, ``Defining and
  classifying software bots: A faceted taxonomy,'' in \emph{Proceedings of the
  1st International Workshop on Bots in Software Engineering}.\hskip 1em plus
  0.5em minus 0.4em\relax IEEE Press, 2019, p. 1–6.

\bibitem{Storey:Dagstuhl:20}
M.-A. Storey, A.~Serebrenik, C.~P. Ros{\'e}, T.~Zimmermann, and J.~D. Herbsleb,
  ``Botse: Bots in software engineering (dagstuhl seminar 19471),'' in
  \emph{Dagstuhl Reports}, vol.~9, no.~11, 2020.

\bibitem{santhanam:PeerJ:22}
S.~Santhanam, T.~Hecking, A.~Schreiber, and S.~Wagner, ``Bots in software
  engineering: a systematic mapping study,'' \emph{PeerJ Computer Science},
  vol.~8, p. e866, 2022.

\bibitem{Savary:SPE:22}
M.~Savary-Leblanc, L.~Burgue{\~n}o, J.~Cabot, X.~Le~Pallec, and S.~G{\'e}rard,
  ``Software assistants in software engineering: A systematic mapping study,''
  \emph{Software: Practice and Experience}, 2022.

\bibitem{Liping:CS:21}
L.~Zhao, W.~Alhoshan, A.~Ferrari, K.~J. Letsholo, M.~A. Ajagbe, E.-V. Chioasca,
  and R.~T. Batista-Navarro, ``Natural language processing for requirements
  engineering: A systematic mapping study,'' \emph{ACM Comput. Surv.}, vol.~54,
  no.~3, apr 2021.

\bibitem{Felfernig:2013}
A.~Felfernig, G.~Ninaus, H.~Grabner, F.~Reinfrank, L.~Weninger, D.~Pagano, and
  W.~Maalej, \emph{An Overview of Recommender Systems in Requirements
  Engineering}.\hskip 1em plus 0.5em minus 0.4em\relax Berlin, Heidelberg:
  Springer Berlin Heidelberg, 2013, pp. 315--332.

\bibitem{hassan:emse:2018}
S.~Hassan, C.~Tantithamthavorn, C.-P. Bezemer, and A.~E. Hassan, ``Studying the
  dialogue between users and developers of free apps in the google play
  store,'' \emph{Empirical Software Engineering}, vol.~23, pp. 1275--1312,
  2018.

\bibitem{Martens:RE:19}
D.~Martens and W.~Maalej, ``Extracting and analyzing context information in
  user-support conversations on twitter,'' in \emph{2019 IEEE 27th
  International Requirements Engineering Conference (RE)}, 2019.

\bibitem{Josephs:ICSA:22}
A.~Josephs, F.~Gilson, and M.~Galster, ``Towards automatic classification of
  design decisions from developer conversations,'' in \emph{2022 IEEE 19th
  International Conference on Software Architecture Companion}, 2022.

\bibitem{Phaithoon:ASE:22}
S.~Phaithoon, S.~Wongnil, P.~Pussawong, M.~Choetkiertikul, C.~Ragkhitwetsagul,
  T.~Sunetnanta, R.~Maipradit, H.~Hata, and K.~Matsumoto, ``Fixme: A github bot
  for detecting and monitoring on-hold self-admitted technical debt,'' in
  \emph{2021 36th IEEE/ACM International Conference on Automated Software
  Engineering (ASE)}, 2021, pp. 1257--1261.

\end{thebibliography}


\end{document}